\documentclass{aamas2017}

\usepackage{subcaption}
\usepackage{color}
\usepackage{microtype}

\makeatletter
\let\@copyrightspace\relax
\makeatother

\begin{document}


\title{Analysing Congestion Problems in Multi-agent Reinforcement Learning \titlenote{This paper expands our AAMAS 2017 extended abstract~\protect{\cite{radulescu2016}} with a detailed description of the RND problem domain, extensive analysis of the resource abstraction method, and additional analysis of the experimental results.}}


\numberofauthors{3}

\author{
\alignauthor Roxana R\u{a}dulescu\\
\affaddr{Artificial Intelligence Lab}\\
   \affaddr{Vrije Universiteit Brussel}\\    
    \affaddr{Belgium}
       \email{rradules@vub.ac.be}
\alignauthor Peter Vrancx\\
\affaddr{Artificial Intelligence Lab}\\
   \affaddr{Vrije Universiteit Brussel}\\    
    \affaddr{Belgium}
       \email{pvrancx@vub.ac.be}
\alignauthor 
Ann Now\'{e}\\
\affaddr{Artificial Intelligence Lab}\\
   \affaddr{Vrije Universiteit Brussel}\\    
    \affaddr{Belgium}
       \email{anowe@vub.ac.be}
}


\maketitle

\begin{abstract}
Congestion problems are omnipresent in today's complex networks and represent a challenge in many research domains. In the context of Multi-agent Reinforcement Learning (MARL), approaches like difference rewards and resource abstraction have shown promising results in tackling such problems. Resource abstraction was shown to be an ideal candidate for solving large-scale resource allocation problems in a fully decentralized manner. However, its performance and applicability strongly depends on some, until now, undocumented assumptions. 
Two of the main congestion benchmark problems considered in the literature are: the Beach Problem Domain and the Traffic Lane Domain. In both settings the highest system utility is achieved when overcrowding one resource and keeping the rest at optimum capacity. We analyse how abstract grouping can promote this behaviour and how feasible it is to apply this approach in a real-world domain (i.e., what assumptions need to be satisfied and what knowledge is necessary). We introduce a new test problem, the Road Network Domain (RND), where the resources are no longer independent, but rather part of a network (e.g., road network), thus choosing one path will also impact the load on other paths having common road segments. We demonstrate the application of state-of-the-art MARL methods for this new congestion model and analyse their performance. RND allows us to highlight an important limitation of resource abstraction and show that the difference rewards approach manages to better capture and inform the agents about the dynamics of the environment.

\end{abstract}


%

\ccsdesc[500]{Computing methodologies~Multi-agent systems}
\ccsdesc[500]{Computing methodologies~Multi-agent reinforcement learning}

\printccsdesc


\keywords{Multi-agent reinforcement learning; Congestion problems; Resource abstraction}

\section{Introduction}
Solving congestion problems is an important research area, as they are present in a wide variety of domains such as traffic control \cite{wiering2000multi}, air traffic management \cite{agogino2012multiagent}, data routing \cite{vicisano1998tcp} and  sensor networks \cite{Wan2003}. Multi-agent reinforcement learning (MARL) has proven to be a suitable framework for such problems \cite{tumer2008aligning, devlin2014potential, malialis2016resource, colby2016local}, as it allows autonomous agents to learn in a decentralized manner, by interacting within a common environment. 

We are interested here in MARL scenarios of independent learners, where no direct communication takes place between the agents. As the task at hand is a resource selection problem where the number of agents exceeds the available capacity, the agents should coordinate in order to achieve a high system utility. We consider here two main approaches designed to achieve this goal: difference rewards \cite{wolpert2001optimal} and resource abstraction \cite{malialis2016resource}. 

Difference rewards is a reward shaping approach that aims to align the interests of agents and the system, as well as to tackle the credit assignment problem in MARL, by informing each agent of its own contribution to the performance of the system. Resource abstraction is a recent approach designed to offer a more informative reward signal that improves learning speed, coordination between agents, as well as the final solution quality. In [8] the authors have shown to outperform difference rewards, however as we show in Section~\ref{sec:exp}, we cannot confirm this conclusion. Our first contribution is that we provide clear insights on how resource abstraction guides the collective behaviour of the agents and we highlight the method's limitations, assumptions and application guidelines.

Current benchmark congestion problems present in the literature often make unrealistic assumptions regarding the independence between the available resources. In complex network management domains, such as smart grids and traffic networks, resources are connected and interdependent, such that using one resource impacts the load of others as well. For this purpose we introduce the Road Network Domain (RND), a problem that models the resources as a system of interconnected roads. We proceed to demonstrate the application of state-of-the-art MARL methods on this problem and analyse their capacity of capturing the newly introduced dynamics in the environment.

The rest of the paper is organized as follows: Section~\ref{sec:back} presents an overview of the theoretical concepts concerning the congestion problem and describes the two main considered resource selection tasks, Section~\ref{sec:RAdiscussion} offers an in depth explanation of the recently introduced resource abstraction method, Section~\ref{sec:RND} introduces a new congestion problem and how to apply existing methods for solving it, Section~\ref{sec:exp} presents and discusses the performed experiments and results, and finally Section~\ref{sec:conc} offers some concluding remarks and future possible directions.

\section{Background}
\label{sec:back}
\subsection{Reinforcement Learning}
Reinforcement Learning (RL) \cite{sutton1998reinforcement} is a machine learning approach which allows an agent to learn how to solve a task by interacting with the environment, given feedback in the form of a reward signal. The solution consists in finding a policy, i.e., a mapping between states and actions that maximizes the received rewards. Q-learning is a common RL value-based method, in which a value function is iteratively updated to optimize the expected long-term reward. After a transition from environment state $s$ to $s'$, through action $a$, Q-learning performs the following update on its estimated state-action value function $Q$, which represents the quality of actions in given states:
\begin{equation*}
Q(s, a) = Q(s, a) + \alpha [r + \gamma \max_{a'}Q(s', a') - Q(s, a)]
\end{equation*}
where $\alpha$ is the learning rate, $\gamma$ is the discount factor and $r$ is the immediate reward received from the environment. In order to address the exploration-exploitation trade-off challenge in RL, one can use the $\epsilon$-greedy action selection method, which allows the agent to choose exploratory random actions with a probability $\epsilon$.

When transitioning to the multi-agent case, we consider  the scenario of independent learners interacting in the same environment. Solving a congestion problem can then be viewed from two perspectives -- agent-centred and system-centred -- that are often in conflict. Allowing selfish entities to act in their own interest in a resource-sharing system can lead to a tragedy of the commons situation \cite{Hardin1243}, which is a detrimental outcome for both the system and the agents.

In multi-agent reinforcement learning a central concern is thus providing a reward signal that will offer a beneficial collective behaviour at the system level. A first approach is providing a \emph{local reward} ($L$) which reflects information about the parts of the system the agent is involved in. $L$ is individually received by each agent and encourages a selfish behaviour, as agents try to optimize their own reward. An alternative approach is the  \emph{global reward} ($G$) which reflects the global system utility and should stimulate agents to perform actions beneficial for the system. The global reward signal can incorporate a significant amount of noise, as the individual effect of an agent on the system's utility can be overshadowed by other learners' effect, i.e., credit assignment problem. Additionally, in large systems, aggregating at each time-step over all the components can be more costly than relying on local information for the reward computation.
\subsection{Resource Selection Congestion Problems}
\label{sec:pb}
A congestion problem from a multi-agent learning perspective is defined by a set of $n$ available resources $\Psi = \{\psi_1,...,\psi_n\}$. Each resource $\psi$ is defined by three properties: $\psi = \langle w_{\psi},c_{\psi},x_{\psi,t}\rangle$, where $w_{\psi} \ge 0$ represents the weighting of the resource, $c_{\psi} > 0$ is the capacity of  $\psi$ and finally $x_{\psi,t} \ge 0$ is the consumption of $\psi$ at time $t$. A resource $\psi$ is congested when $x_{\psi,t} > c_{\psi}$. The local utility of a resource $\psi$ is defined in terms of its properties:
\begin{equation}
L(\psi,t) = f(x_{\psi,t},c_{\psi},w_{\psi})
\label{eq:genLR}
\end{equation}

\begin{figure*}[!t]
\centering
\begin{subfigure}{.33\textwidth}
  \centering
  \includegraphics[width=\linewidth]{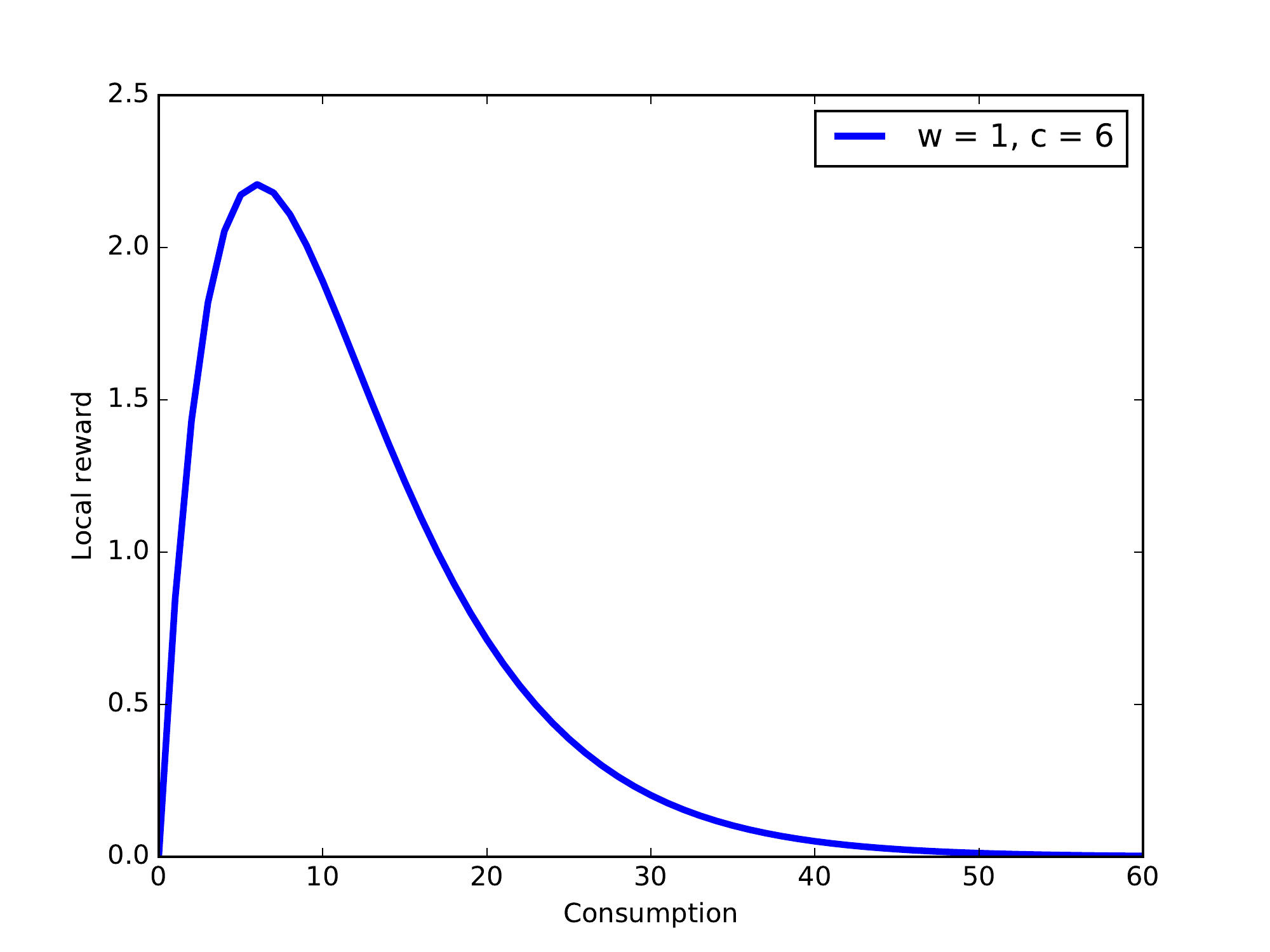}
  \caption{Local reward BPD, Equation~\ref{eq:bpd}}
  \label{fig:beachR}
\end{subfigure}%
\begin{subfigure}{.33\textwidth}
  \centering
  \includegraphics[width=\linewidth]{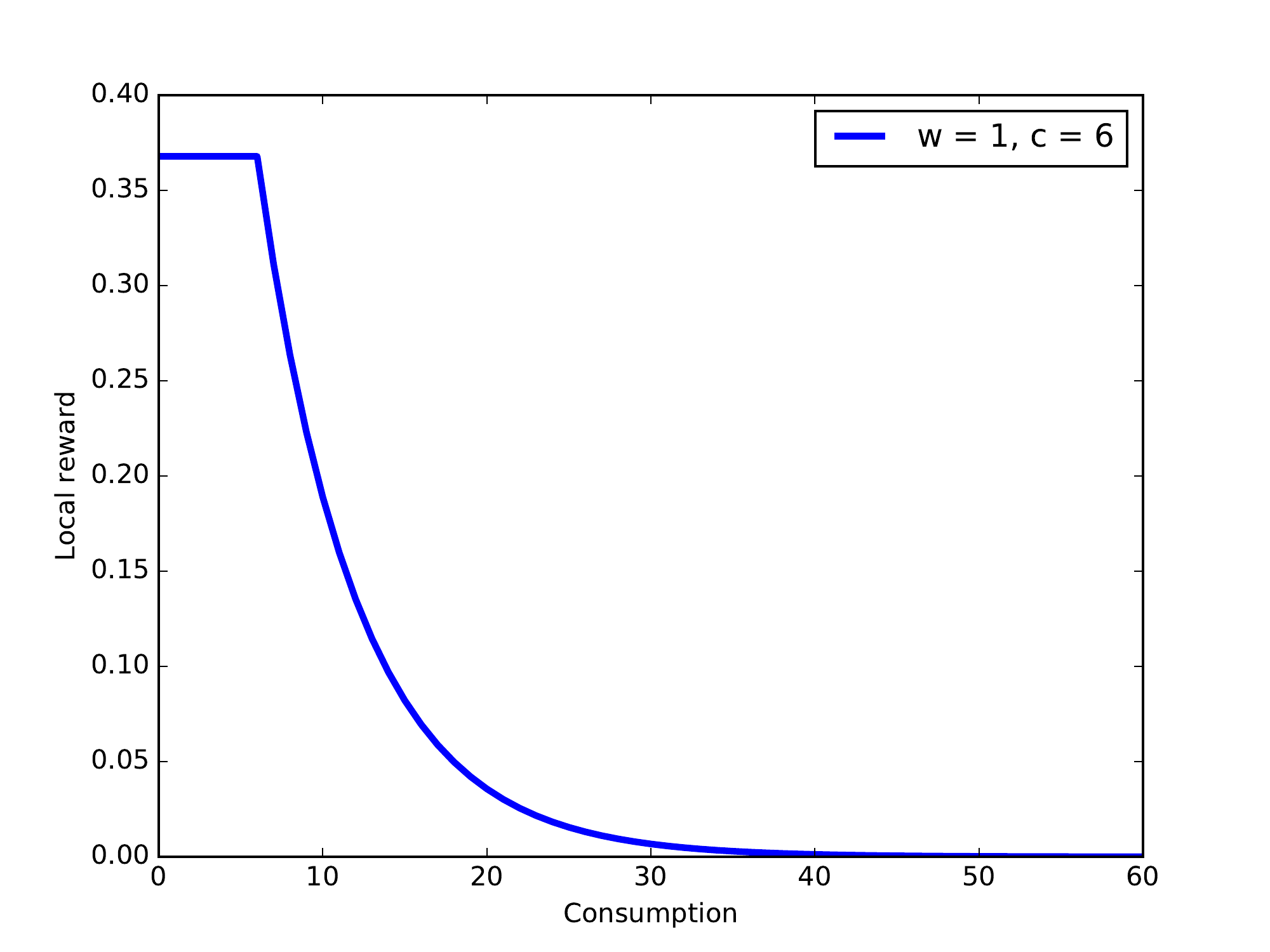}
  \caption{Local reward TLD, Equation~\ref{eq:tld}}
  \label{fig:trafficR}
\end{subfigure}%
\begin{subfigure}{.33\linewidth}
\centering
  \includegraphics[width=\linewidth]{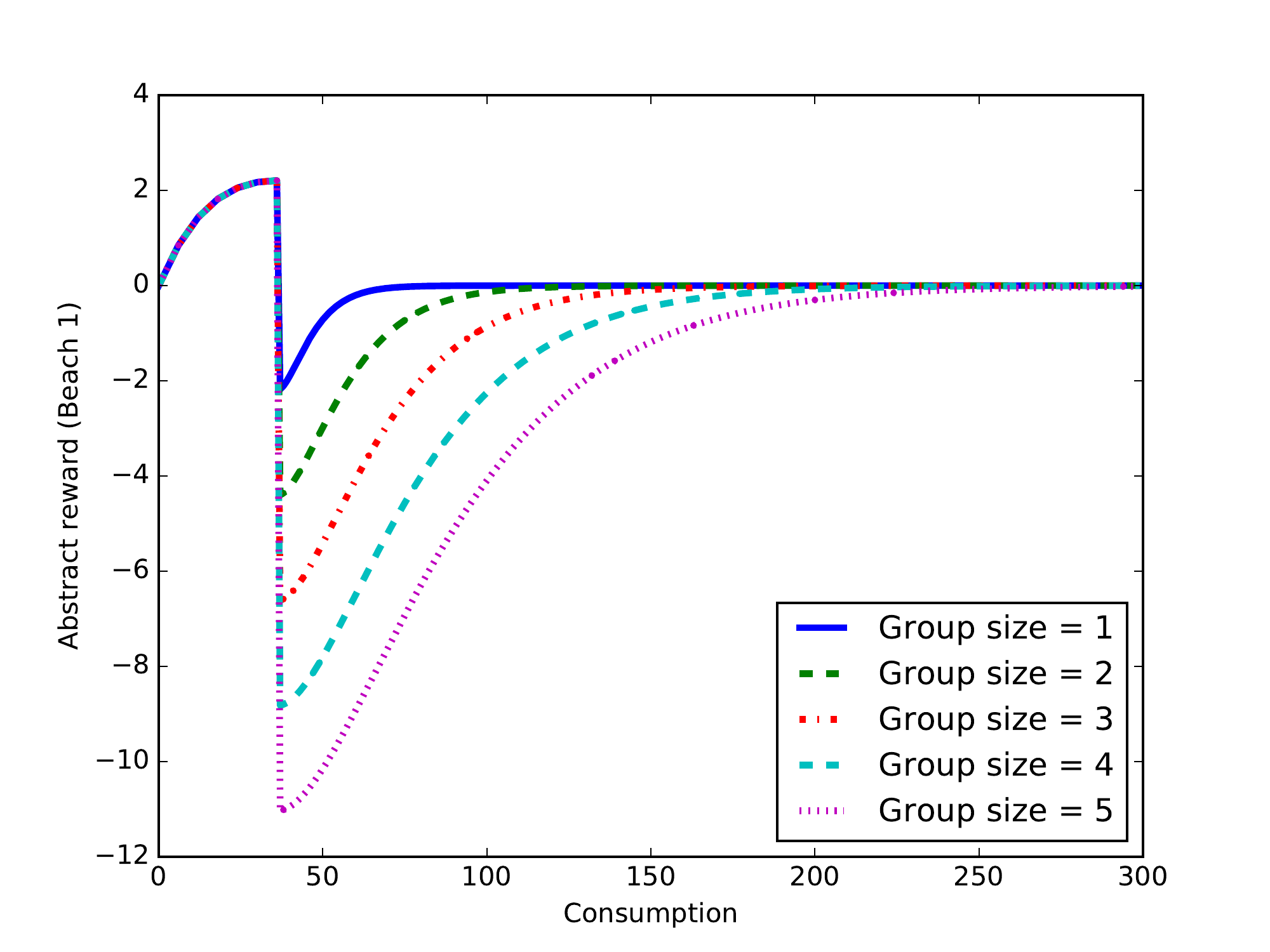}
  \caption{Abstract reward BPD, Equation~\ref{eq:abstractR}}
  \label{fig:raR}
\end{subfigure}
\caption{Local reward (TLD and BDP) and abstract reward (BDP) received by the agents selecting section 1 from 5 available sections, each with capacity 6 and weight 1. After the congestion point of the resource is reached, the abstract reward approach reflects the reward function over the x-axis, giving agents a high initial penalty for starting to overcrowd a section.}
\label{fig:localR}
\end{figure*}

In this paper we consider two resource selection problems that have become benchmark problems for studying resource allocation in RL. They mainly differ with respect to their local utility schemes. The first problem is defined as the \emph{beach problem domain} (BPD) \cite{tumer2013coordinating}, where all the available resources are considered beach sections with the same weight equal to 1 and the same capacity $c$:
\begin{equation}
L(\psi,t) = x_{\psi,t} e^\frac{-x_{\psi,t}}{c}
\label{eq:bpd}
\end{equation}
The second type of problem is the \emph{traffic lane domain} (TLD) \cite{tumer2009traffic}, where the agents have to select between several available lanes, each having a different capacity and weight (reflecting the importance or desirability of the lane):
\begin{align}
L(\psi,t) &= 
\begin{cases}
w_{\psi} e^{-1}& ,x_{\psi,t}\le c_{\psi} \\
w_{\psi} e^\frac{-x_{\psi,t}}{c_{\psi}} & ,x_{\psi,t}> c_{\psi}
\end{cases}
\label{eq:tld}
\end{align}
For both problems the global utility is defined as the sum over all the local utilities at time $t$:
\begin{equation}
G(t)= \sum_{\psi \in \Psi}L(\psi,t)
\label{eq:globalR}
\end{equation}
At each time step the agents choose to move to a certain beach section or traffic lane and receive a reinforcement in accordance to the effect of their joint action on the implemented reward scheme. The main difference between the two described local utility functions is represented by the segment before the congestion point is reached. For BPD (Figure~\ref{fig:beachR}) the maximum utility for a resource is achieved at optimum capacity ($x_{\psi,t}= c_{\psi}$), while for TLD (Figure~\ref{fig:trafficR}) this condition is less strict, only requiring the lanes to be under the congestion point ($x_{\psi,t} \le c_{\psi}$).

If the number of agents exceeds the total capacity, the configuration achieving the \emph{highest global utility} for these benchmark problems is one that overcrowds one of the resources and leaves the rest at optimum capacity. For the BPD the congested resource can be any of the available beach sections, while for the TLD it should be the lane with the lowest weight and highest capacity combination \cite{devlin2014potential}. 

\subsection{Difference Rewards}
Driven by the idea that the reward signals should allow the agents to deduce their individual contribution to the system, a reward signal we consider here is \emph{difference rewards} ($D$)~\cite{wolpert2001optimal}. Under a global system utility $G$, the difference rewards for agent $i$ is defined as:
\begin{equation}
D_i(z) = G(z) - G(z_{-i})
\label{eq:dr}
\end{equation}
where $z$ denotes a general term for either state, action or state-action pair, according to the considered domain, and $G(z_{-i})$ is the global utility of a system from which the contribution of agent $i$ is removed. The difference rewards signal was constructed following two main guidelines: (i) aligning the reward received by the agent with the global system utility -- \emph{factoredness}, while (ii) lowering the impact of other agents on the reward signal -- \emph{learnability}, thus addressing the credit assignment problem in a multi-agent setting  \cite{agogino2008analyzing, tumer2013coordinating}. These characteristics have proven to significantly improve learning performance \cite{tumer2009traffic, malialis2014intrusion, tumer2008aligning}, even in domains where $G(z_{-i})$ cannot be directly computed and should be estimated \cite{agogino2012multiagent, Colby2013, colby2016local}.  

We now take a look at how the presented credit assignment approach can be defined for these congestion models. By combining Equations~\ref{eq:bpd}, \ref{eq:globalR} and \ref{eq:dr} we obtain the following formulation of the difference rewards for the BPD case:
\begin{equation}
D_i(t)=L(\psi,t) - (x_{\psi,t} -1)e^\frac{-(x_{\psi,t}-1)}{c}
\label{eq:drBPD}
\end{equation} 
as the impact of agent $i$ is solely limited to the chosen resource $\psi$, thus all the other terms in the sum composing the global reward cancel out. The same approach can be used for Equations~\ref{eq:tld} and \ref{eq:dr} for the TLD case.
\subsection{Resource Abstraction}
\label{sec:RAdef}
\emph{Resource abstraction} \cite{malialis2016resource} is an approach that aims to improve learning speed, solution quality and scalability in large MARL  congestion problems. Resource abstraction provides the agents with a more informative reward, facilitating coordination in systems with up to 1000 agents \cite{malialis2016resource}.

Resource abstraction entails grouping a set of resources into disjoint subsets, and modifying the local reward function after the congestion point of a resource is reached, such that agents using it will get a higher penalty for overcrowding the resource.

An abstract group is defined by aggregating the properties of the composing resources. In the congestion model defined above, an abstract group $b$ has the following properties: consumption $X_{b,t}=\sum_{\psi \in b}x_{\psi,t}$, capacity  $C_b=\sum_{\psi \in b}c_{\psi}$ and weight $W_b=\frac{1}{|b|}\sum_{\psi \in b}w_{\psi}$. In other words, resource abstraction creates virtual resources which are agglomerations of resources. Given a resource $\psi$ and the abstract group $b$ to which it belongs to, the abstract reward for an agents using resource $\psi$ at time $t$ for the BPD is defined as:
\begin{align}
A(b,\psi,t) &= 
\begin{cases}
L(\psi,t), & x_{\psi,t}\le c_{\psi} \\
- X_{b,t}e^\frac{-X_{b,t}}{C_b},& x_{\psi,t}> c_{\psi} 
\end{cases}
\label{eq:abstractR}
\end{align}
The same approach can be used for the TLD case. 

\section{Analysing Resource Abstraction}
\label{sec:RAdiscussion}
Notice that, in order to apply resource abstraction, information is required regarding the weight, capacity and consumption of the resources as well as the system utility function, limiting the straightforward applicability on a real-world domain, where such information might not be available. Even though the work \cite{malialis2016resource} presents a few guidelines and remarks on how the resource abstraction should be applied,  clear insights and explanations on how to create the group abstractions are not present. We consider that a more thorough understanding of the method is beneficial to extend its usability and applicability.

For a better understanding of how the resource abstraction impacts the collective behaviour of the agents, we turn to Figure~\ref{fig:raR}. We plot the abstract reward function of various groups in the BPD, differing in the number of composing resources (ranging from size 1 to size 5). There are 5 available beach sections, each with weight equal to 1 and capacity equal to 6, giving a total capacity of 30 for the entire beach. We fill the resources uniformly until we reach the maximum capacity, after which we proceed to overcrowd each of the abstract groups (by overcrowding one of the composing resources). 

We first remark that after the congestion point of a  resource is reached, the abstract reward mirrors the reward function over the x-axis (this is due to the second case of the function presented in Equation~\ref{eq:abstractR}). This causes the initial penalty for starting to overcrowd a section to be more severe (i.e., a negative reward) compared to the local reward presented in Figure~\ref{fig:beachR}. However, continuing to overcrowd a section results in smaller and smaller penalties and given enough agents it will eventually converge to 0. 

A second remark can be made on the effect of the group size. The initial penalty for congesting a resource is correlated with the size of the group, thus we expect agents to prefer overcrowding resources that are part of the smallest abstract groups (e.g., in the case presented in Figure~\ref{fig:raR} starting to overcrowd a group of size 1 returns a reward of -2, while for a group of size 5 the reward is around -11). The same reasoning can be applied for the TLD. 

Lastly, we note that in order to determine the best configuration for the abstract groups, one should also have knowledge on the final desired collective behaviour of the agents (e.g., for the BDP having a group of size 1 should lead to the desired "overcrowd one" behaviour). For more complex domains, finding the optimal abstracts grouping can prove to be impossible, as we show in Section~\ref{sec:exp}, due to the fact that the resource abstraction approach can no longer capture the required collective behaviour.

\section{Road Network Domain}
\label{sec:RND}
\begin{figure}[!h]
\centering
\includegraphics{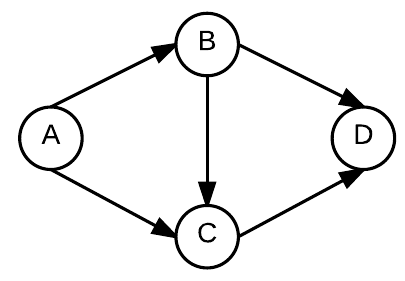}
\caption{Topology example for the RND problem. The agents have to travel from point $A$ to point $D$ and have 3 possible paths: $ABD$, $ABCD$, or $ACD$.}
\label{fig:Braess}
\end{figure}

We propose the \emph{Road Network Domain} (RND), a problem that introduces a scenario in which the resources are not independent, as using one path introduces additional load for others as well. We model this problem as a network of roads (e.g., Figure~\ref{fig:Braess}), where the agents have to choose between paths of the network. Figure~\ref{fig:Braess} presents an example of a small network topology that can be explored, but that already serves the points we want to make.  Each road segment is modelled as a resource, corresponding to the description presented in Section~\ref{sec:pb}. The RND can be used with the utility functions of both BPD and TLD, with the former creating a more challenging task, as the maximum value of the utility is only achieved at optimum capacity. The local reward of a path $P$ is then simply the sum over all the local rewards of the composing road segments $\psi$ (e.g., roads $AB$ and $BD$ for the path $ABD$):
\begin{equation}
L_{path}(P,t) = \sum_{\psi \in P}{L(\psi,t)}
\end{equation}
We compute the global system utility by summing over all the local rewards of the roads segments present in the network (see Equation~\ref{eq:globalR}).\footnote{We sum over road segments rather than over paths in order to avoid having segments that belong to multiple path contributing more than once to the global utility value.}

We consider that the RND introduces a challenge that is often present in real-world domains (e.g., electricity grids, traffic networks). Additionally, one can always increase the difficulty of the problem by creating more complex network structures and can easily translate in this model any real-world situation of interest.

\subsection{Difference Rewards}
As the impact of agent $i$ on the system is limited to the composing road segments of his chosen path $P$, we can define the difference rewards for the RND as follows, where $f$ is a local reward function (see Equation~\ref{eq:genLR}):
\begin{equation}
\begin{split}
D_i(t) & = L_{path}(P,t) - L_{path}(P_{-i},t) \\
& = L_{path}(P,t) - \sum_{\psi \in P} f(x_{\psi,t}-1,c_{\psi},w_{\psi})
\end{split}
\end{equation} 
\subsection{Resource Abstraction}
We consider here two approaches for defining the abstract group construction for the resource abstraction method: over \emph{road segments} or over \emph{paths of the network}. As a road segment is a resource, the properties of an abstract group over a set of segments coincide with the ones defined in section~\ref{sec:RAdef}. The abstract reward for each road segment $\psi$ and its corresponding group $b$ is defined as:
\begin{align}
A(b,\psi,t) &= 
\begin{cases}
L(\psi,t), & \ x_{\psi,t}\le c_{\psi}\\
- f(X_{b,t},C_b,W_b), & x_{\psi,t}> c_{\psi}
\end{cases}
\label{eq:abstractsegRND}
\end{align}
with $f$ a local reward function (see Equation~\ref{eq:genLR}).
Finally, the abstract reward for choosing a path $P$ at time $t$ then becomes the sum over the abstract reward of each composing road segment:
\begin{align}
A_{path}(P,t) &= \sum_{\psi \in P}{A(b,\psi,t)}
\label{eq:abstractsegmRND}
\end{align}

We also extend the definition for an abstract group $b$ over a set of paths: consumption $X_{b,t}=\sum_{P \in b}x_{P,t}$, capacity  $C_b=\sum_{P \in b}c_{P}$ and weight $W_b=\frac{1}{|b|}\sum_{P \in b}w_{P}$, where $x_{P,t}$ is the number of agents that choose path $P$, $c_P=min_{\psi \in P}(c_{\psi})$ and $w_P=\frac{1}{|P|}\sum_{\psi \in P}w_{\psi}$.  We consider a path to be congested if any of its composing roads is congested. We can now define the abstract reward for a selected path $P$ at time $t$ as:
\begin{align}
A(b,P,t) &= 
\begin{cases}
L_{path}(P,t), & \forall \psi  \in P\ :\ x_{\psi,t}\le c_{\psi}\\
- f(X_{b,t},C_b,W_b), & \exists \psi \in P\ :\ x_{\psi,t}> c_{\psi}
\end{cases}
\label{eq:abstractRND}
\end{align}
where $b$ is the corresponding abstract group of $P$.
%

Next, we present a series of experiments designed to demonstrate how to best use the resource abstraction method, but also its limitations. Additionally, we test all the presented approaches (local and global rewards, difference rewards and resource abstraction) on the RND, and evaluate which method can best model the underlying dynamics of the environment in the reward signal.

\section{Experiments}
\label{sec:exp}
Each agent uses the Q-learning algorithm with an exploration parameter $\epsilon=0.05$ and an exploration decay rate of $0.9999$. As an important aspect of work is to understand and explore resource abstraction, the parameters for the experiment in Section~\ref{sec:expBPD} were chosen to match the setting used in the original work~\cite{malialis2016resource}: learning rate $\alpha = 0.1$, decay rate for $\alpha$ is $0.9999$ and the discount factor $\gamma=1.0$. As for the RND experiments (Section~\ref{sec:expRND}), after several tests, the parameters chosen for the local reward $L$, global reward $G$ and difference rewards $D$ are: learning rate $\alpha = 0.1$, no decay, and the discount factor $\gamma=0.9$.
\subsection{Beach Problem Domain}
\label{sec:expBPD}
Our first experiment is performed on the BPD and it aims to explain and demonstrate the use of resource abstraction. We borrow the original setting from \cite{malialis2016resource}, with 6 beach sections, each with capacity 6, thus a total system capacity of 36. There are 100 agents (creating a congestion scenario) and the maximum system utility is 11.04, achieved when overcrowding one of the sections with 70 agents, while keeping the other five at the optimum capacity of 6 agents. Each agent has three available actions: shift to the resource on the left, shift to the one on the right and maintain position and has $5$ time-steps to finalize his action sequence for an episode. We run the scenario for $4\,000$  episodes and plot the global utility averaged over 50 trials, together with error bars representing the standard deviation at every 500 episodes. Recall that the local reward function for the BPD is the one plotted in Figure~\ref{fig:beachR}.
\begin{figure}[!h]
\centering
\includegraphics[clip, trim=1cm 0.3cm 1.3cm 0.9cm,width=\linewidth]{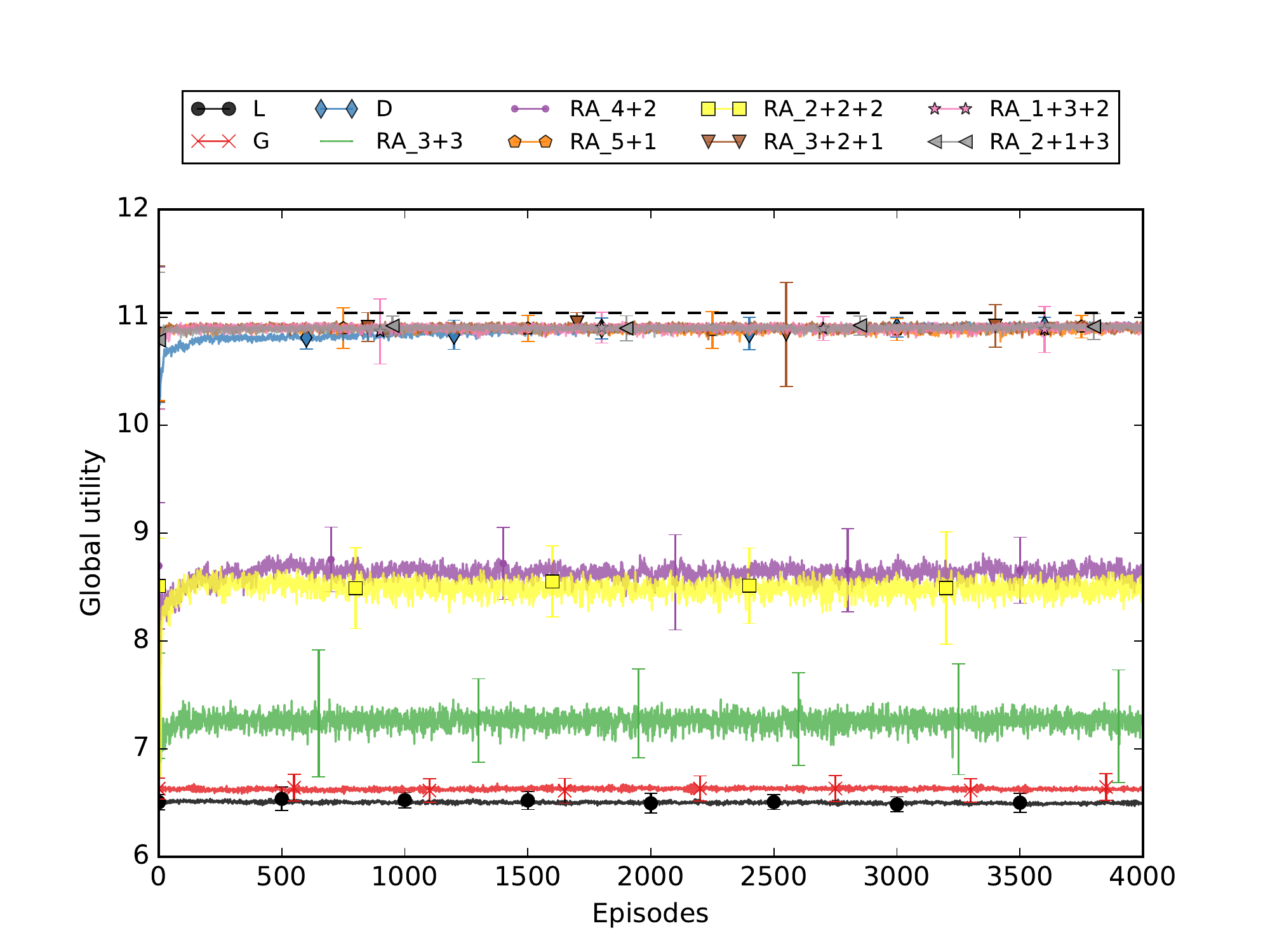}
\caption{BPD, 100 agents, 6 sections each with capacity 6. $D$ together with all the $RA$ configurations containing an abstract group of size 1 achieve the highest performance.}
\label{fig:exp1}
\end{figure}

Figure~\ref{fig:exp1} presents the results obtained for the BPD for the following reward schemes: local reward $L$, global reward $G$, difference rewards $D$ and resource abstraction $RA$. There are seven different resource abstraction configurations, just like in \cite{malialis2016resource}, composed of either two or three abstract groups. For example $RA\_3+2+1$ denotes that the first 3 sections form one abstract group, the next 2 another one and the last section represents an abstract group on its own.

Given the insights presented in the previous section on how the grouping of the resources influences the collective behaviour and knowing that the maximum utility is achieved under the `overcrowd one' behaviour, we expect that the RA with abstract groups containing only one resource should attain the best performances. The results presented in Figure~\ref{fig:exp1} confirm our expectations and provide the following ranking of the $RA$ configurations: all the variations containing a group of one resource achieve the best performance, $RA$ configurations where 2 is the smallest group size come in second, while the grouping $RA\_3+3$ comes in last, but still above the $G$ and $L$ reward schemes schemes.

An important remark we make here concerns the performance of the difference rewards $D$. In our experiment $D$ ranks among the best performing methods, in contrast to the results obtained in \cite{malialis2016resource}, where $D$ plateaued around the value of 8. We also note that the difference rewards application to the BPD described in Equation 14 of the work  \cite{malialis2016resource} does not correspond to the equation we consider to be the correct one (Equation~\ref{eq:drBPD}). \footnote{$D_i(t)=L(\psi,t) - x_{\psi,t} e^\frac{-(x_{\psi,t}-1)}{c}$ \cite{malialis2016resource}, whereas the second term should be a function of $(x_{\psi,t}-1)$.} 

\begin{figure}[!h]
\centering
  \includegraphics[width=.83\linewidth]{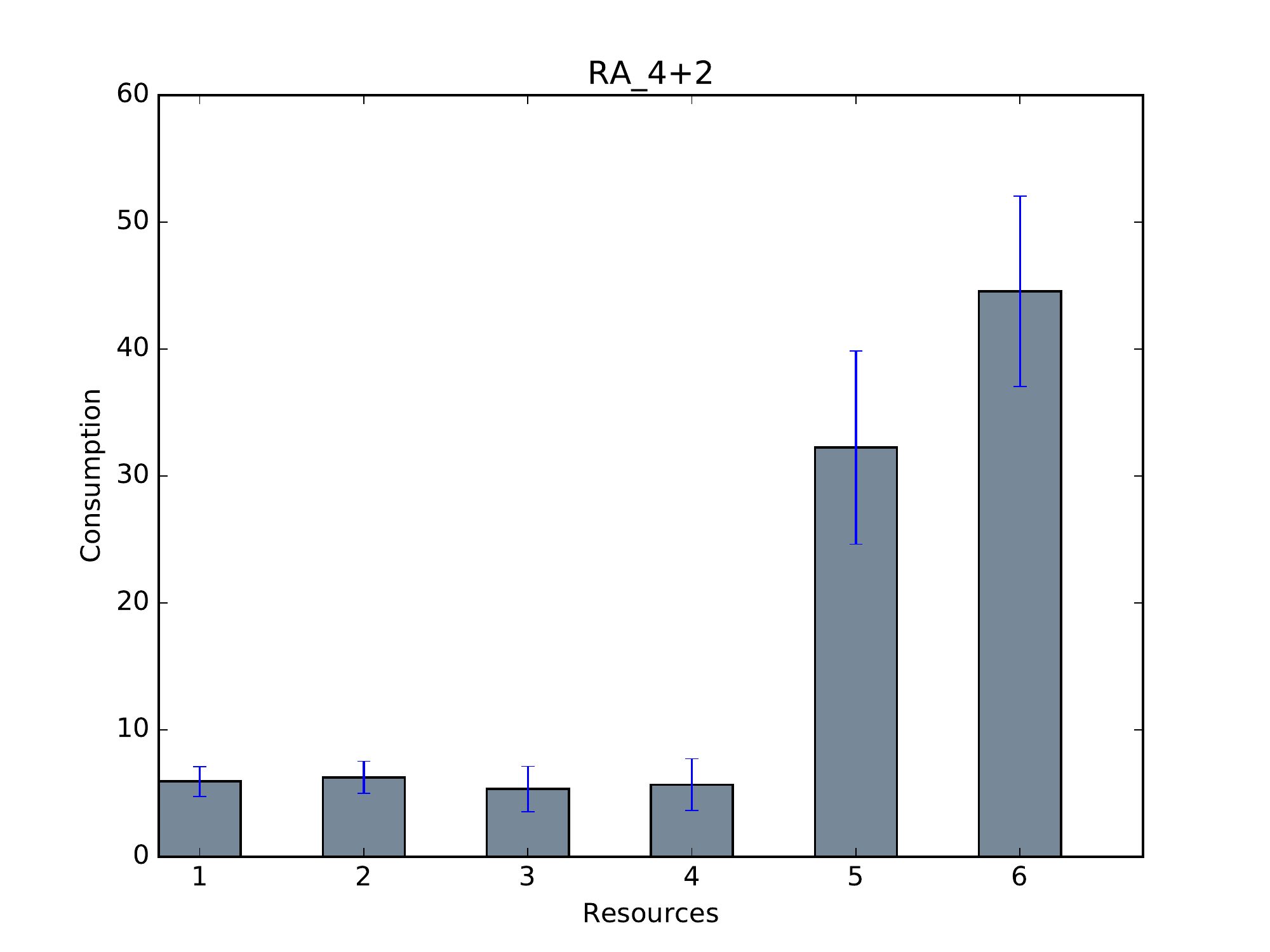}
  \caption{BPD, distribution of agents over the 6 beach sections for the $RA\_4+2$ setting. The agents choose to overcrowd the group with the smallest size, meaning resources 5 and 6.}
  \label{fig:ra42}
\end{figure}

\begin{figure}[!h]
  \centering
  \includegraphics[width=.83\linewidth]{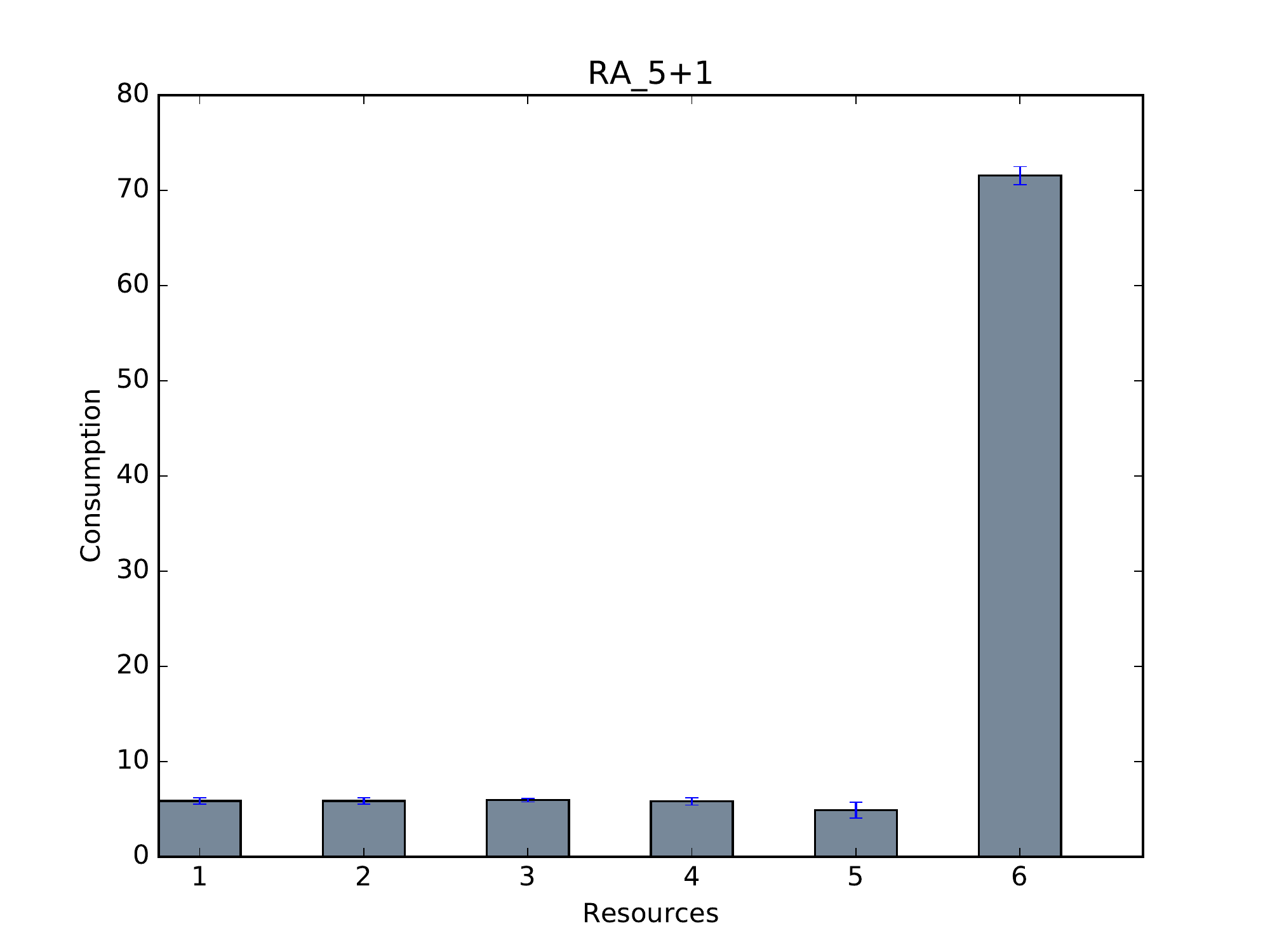}
  \caption{BPD, distribution of agents over the 6 beach sections for the $RA\_5+1$ setting. The agents choose to overcrowd the group with the smallest size, meaning resource 6. This distribution is one of the possible optimum configurations.}
  \label{fig:ra51}
  \end{figure}

\begin{figure}[!h]
  \centering
  \includegraphics[width=.83\linewidth]{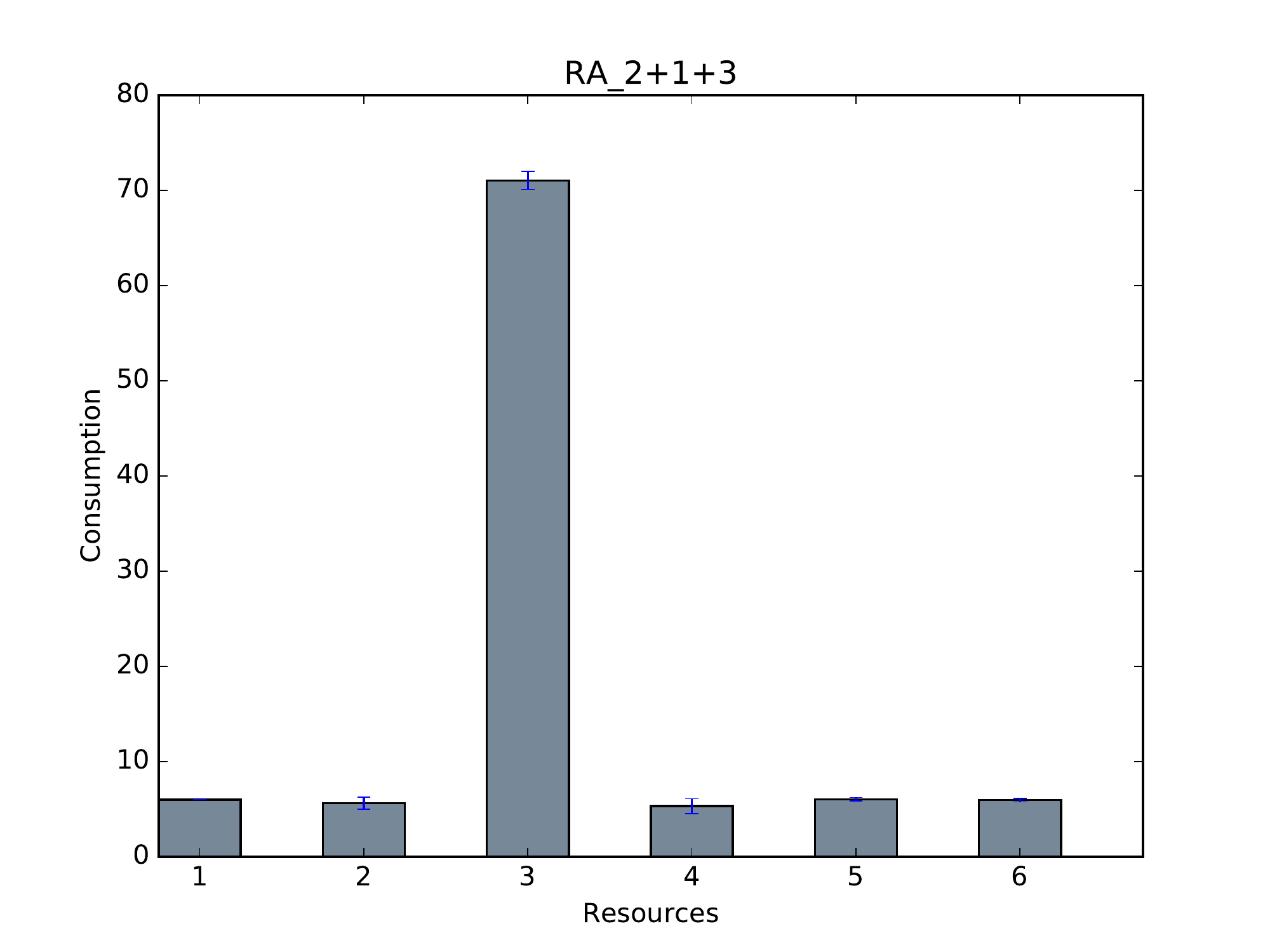}
  \caption{BPD, distribution of agents over the 6 beach sections for the $RA\_2+1+3$ setting. The agents choose to overcrowd the group with the smallest size, meaning resource 3. This distribution is one of the possible optimum configurations.}
  \label{fig:ra213}
\end{figure}

Figures~\ref{fig:ra42}, \ref{fig:ra51}, and \ref{fig:ra213} present the final distributions of agents over the  resources for the configurations $RA\_4+2$, $RA\_5+1$, and $RA\_2+1+3$. The results are averaged over 50 trials and each bar plot is accompanied by the standard deviation error. Notice that in each case the agents overcrowd the resources corresponding to the smallest abstract group. $RA\_5+1$ and $RA\_2+1+3$ present two examples of how the optimal configuration can look like. We note that we did not include the representations for  $RA\_3+2+1$ and  $RA\_1+3+2$ as they are similar to $RA\_2+1+3$. Visualizing the distribution for configurations like $RA\_2+2+2$ or $RA\_3+3$ is not possible, as the agents can end up overcrowding any of the abstract groups.

\subsection{Road Network Domain}
\label{sec:expRND}
For our next experiments we test how the considered reward schemes perform on the newly introduced Road Network Domain. As previously mentioned, RND is designed to work with any of the local utility functions presented in Section~\ref{sec:pb}. We create two types experiments, one using the local utility of BPD and another with TLD. For all the experiments we use the network topology presented in Figure~\ref{fig:Braess}. There are 50 agents, all starting at point $A$ and having to reach point $D$. They can choose between three paths: $ABD$, $ACD$, or $ABCD$. Each $RA$ configuration is expressed using a bracket notation depicting the abstract groups over either paths or road segments. For example, in the setting $[ABD,ACD],[ABCD]$ there are two groups: one of size 2 ($ABD$ and $ACD$) and one of size 1 ($ABCD$).

\begin{figure}[!h]
  \centering
  \includegraphics[width=.5\linewidth]{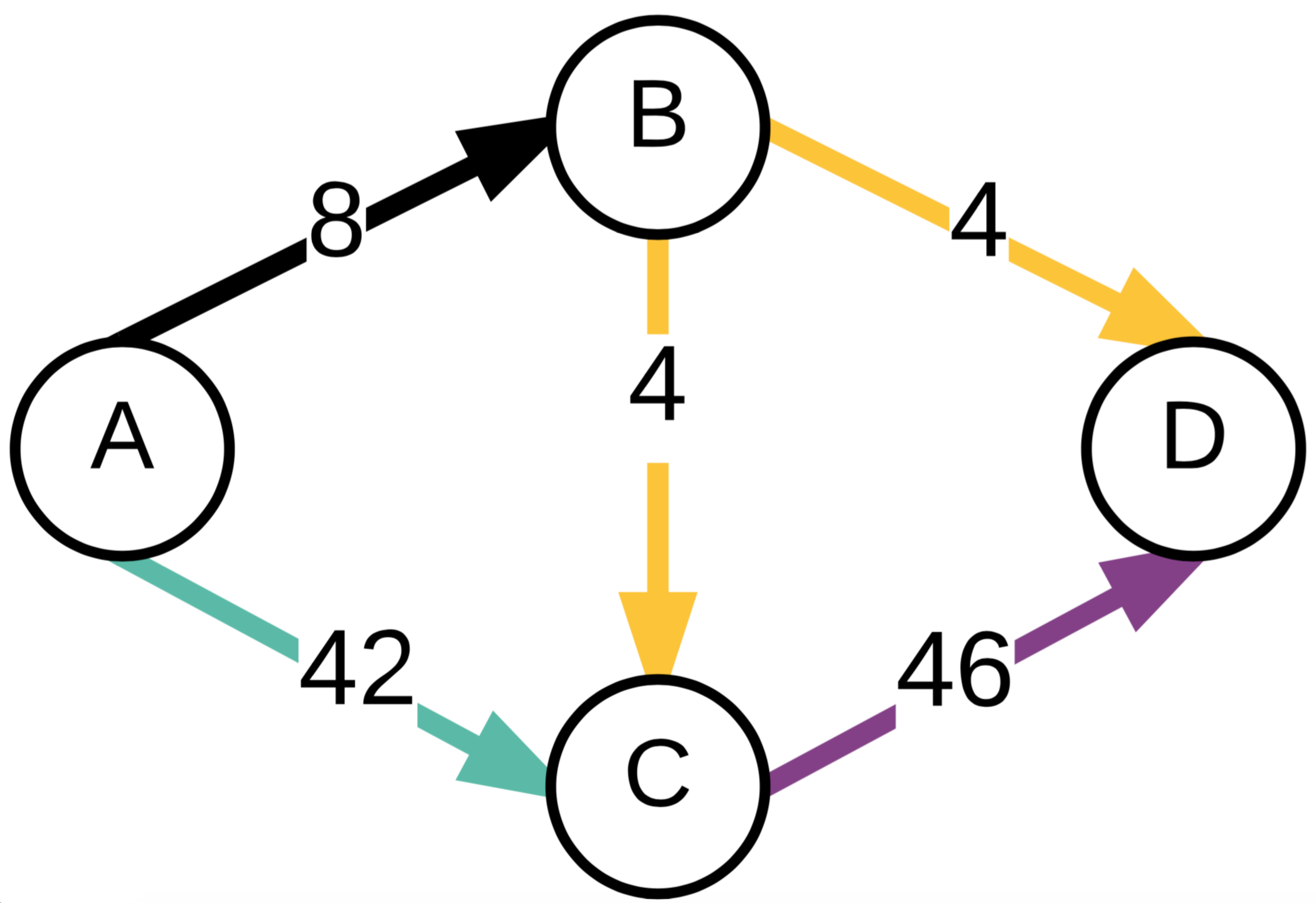}
  \caption{Example of the optimum distribution of 50 agents over the network under the BPD local utility, where each road segment has capacity 5 and weight 1. The colors encode the best performing abstract group configuration over road segments found for this case: $[AB],[AC],[CD],[BC,BD]$.}
  \label{fig:bestRASGM}
 \end{figure}

For the BPD utility case all roads have a capacity of 5 and weight 1. We run the first experiment, using $RA$ over the paths of the network, for $20\,000$  episodes and plot the global utility averaged over 30 trials, together with error bars representing the standard deviation at every $1\,000$ episodes. Figure~\ref{fig:bestRASGM} exemplifies one of the possible optimum configurations. The maximum global utility for this scenario is 5.22.
\begin{figure}[!h]
  \centering
  \includegraphics[clip, trim=1cm 0cm 1.3cm 0cm,width=\linewidth]{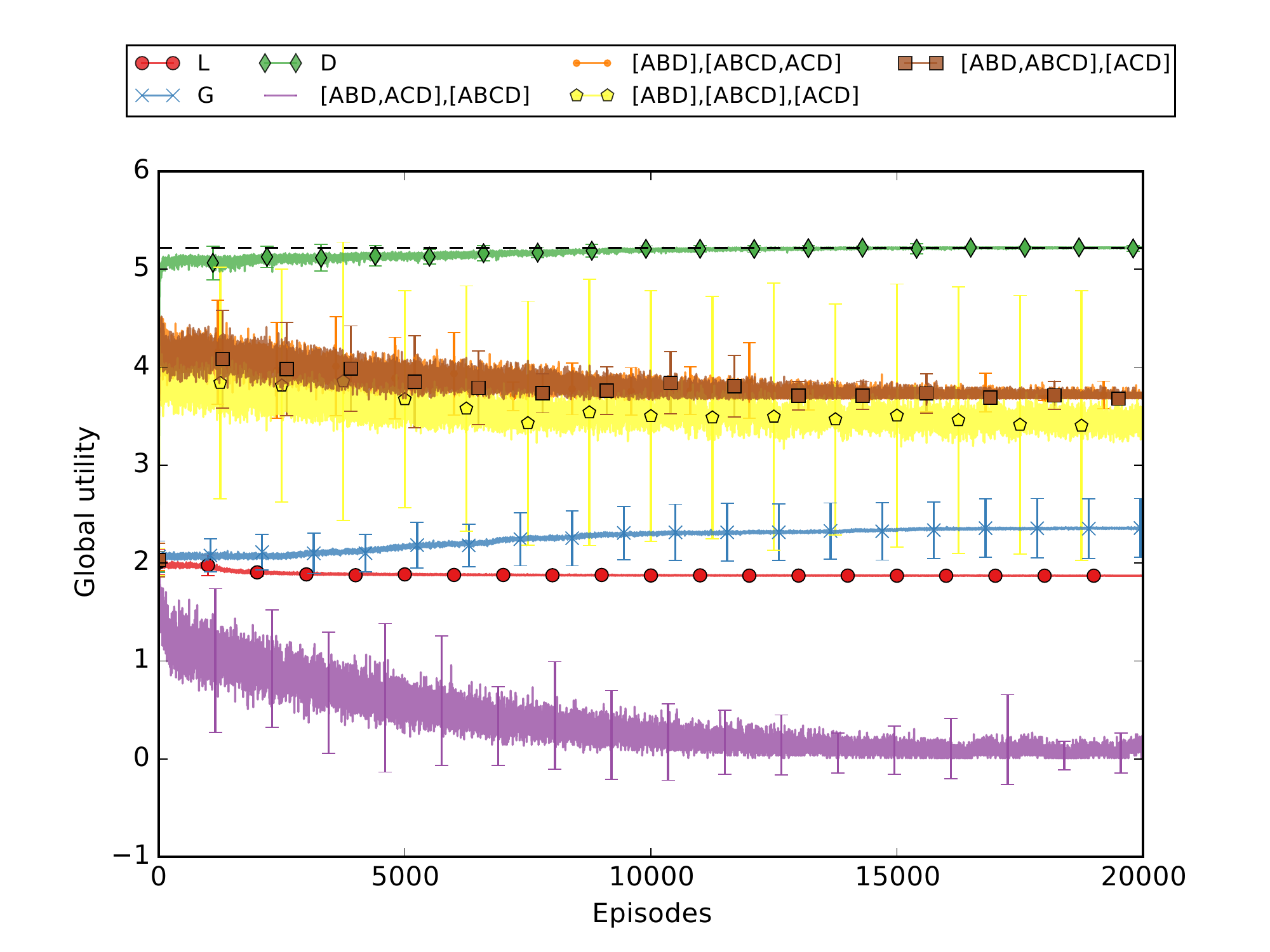}
  \caption{RND with BPD local utility, $RA$ defined over paths, 50 agents. $RA$ can no longer capture the optimal required behaviour, due to the dependencies between resources and shape of the utility function. D manages to reach optimal performance.}
  \label{fig:braessBPD}
 \end{figure}

\begin{figure}[!h]
  \centering
  \includegraphics[clip, trim=1cm 0cm 1.3cm 0cm,width=\linewidth]{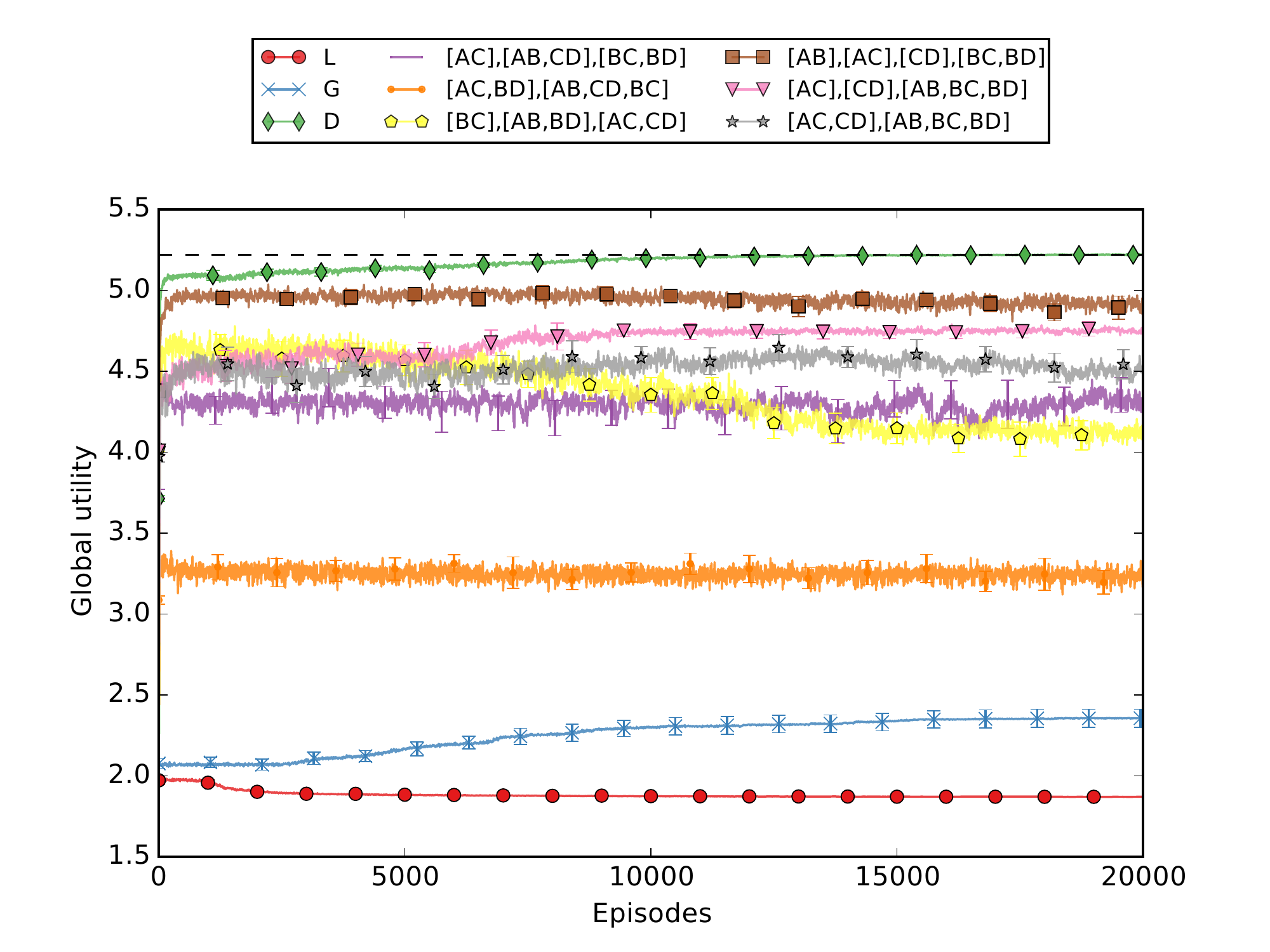}
  \caption{RND with BPD local utility, $RA$ defined over road segments, 50 agents. Even at segment level $RA$ is unable to capture the optimal required behaviour, due to the dependencies between resources and shape of the utility function. D manages again to reach optimal performance.}
  \label{fig:braessBPDSGM}
 \end{figure}
  
The results are presented in Figure~\ref{fig:braessBPD}. We remark that none of the $RA$ settings is able to capture the underlying optimal configuration, as this can no longer be expressed in terms of a pure `overcrowd one' behaviour, due to the dependencies among the resources and the shape of the utility function. This is a situation for which there does not exist an abstract grouping to guide the agents towards the optimal performance, $[ABD, ACD],[ABCD]$ even performs worse than $L$ or $G$, as this setting drives agents to overcrowd the path $ABCD$, contributing to the congestion of the other two paths as well. On the other hand, $D$ manages to achieve the optimal performance in this scenario, demonstrating its capacity to allow agents to adapt to more difficult environment dynamics.
 
We perform a second experiment using the BPD local utility scheme, defining the $RA$ over road segments this time, in order to verify whether at this resource granularity level we can achieve an optimum group abstraction. Notice that, in this case, the number of possible $RA$ settings is much higher, making the decision about the abstract group creation even harder. The $L$, $G$, and $D$ results are the same as in Figure~\ref{fig:braessBPD}. We note that additional smoothing was performed, to improve the visibility of the results obtained for the $RA$ and each plot is accompanied by the standard error. The results of this experiment are presented in Figure~\ref{fig:braessBPDSGM}. We remark again that none of the RA settings manage to reach the optimum solution, with the best performing one being $[AB],[AC],[CD],[BC,BD]$. A visual representation of this best performing grouping can be found in Figure~\ref{fig:bestRASGM}: the two segments that need to be kept under the congestion point ($BC,BD$) form the largest group, while all the others form their own abstract group. Thus, the RND problem has allowed us to demonstrate that having disjoint abstract groupings is not a sufficient condition for being able to reach an optimum solution and that the necessity of having independent resources goes beyond having segments not belonging to the same abstract group.
 
To better understand these results, we can turn again to Figure~\ref{fig:bestRASGM}. Notice that even though the capacity of the road segments is 5, the optimum configuration does not include any segments having reached this value. We conclude that we cannot express the solution as `overcrowd these segments and keep the rest at optimum capacity', thus being unable to properly express the desired solution using the $RA$ approach.

\begin{figure}[!h]
  \centering
  \includegraphics[width=.65\linewidth]{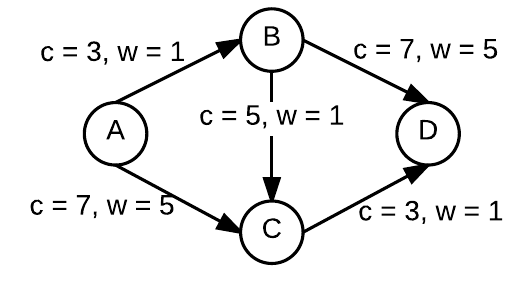}
  \caption{Weights and capacities for each road segment in the RND scenario considered under the TLD local utility.}
  \label{fig:RNDT}
\end{figure}

\begin{figure}[!h]
  \centering
  \includegraphics[width=.5\linewidth]{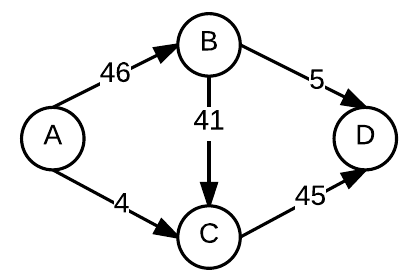}
  \caption{Example of the optimum distribution of 50 agents over the considered RND under the TLD local utility}
  \label{fig:optRNDT}
\end{figure}

For the TLD utility case we use the network scenario described in Figure~\ref{fig:RNDT}. Increasing the weights for $AC$ and $BD$ determines the maximum global utility to be achieved when avoiding to overcrowd these segments and their corresponding paths. Additionally, because the condition for receiving the highest utility for a road segment is less strict compared to the BPD utility scheme (maintaining the consumption under the congestion point versus reaching the optimum capacity), the maximum global utility is achieved when overcrowding the path $ABCD$. An example of such a configuration is presented in Figure~\ref{fig:optRNDT}. We thus expect $RA$ over paths to display a good performance in this scenario, as it should be able to capture the desired behaviour. We run this experiment for $2\,000$  episodes and plot the global utility averaged over 30 trials, together with error bars representing the standard deviation at every $100$ episodes. The highest global utility is 3.68. 

\begin{figure}[!h]
  \centering
  \includegraphics[clip, trim=1cm 0cm 1.5cm 0cm,width=\linewidth]{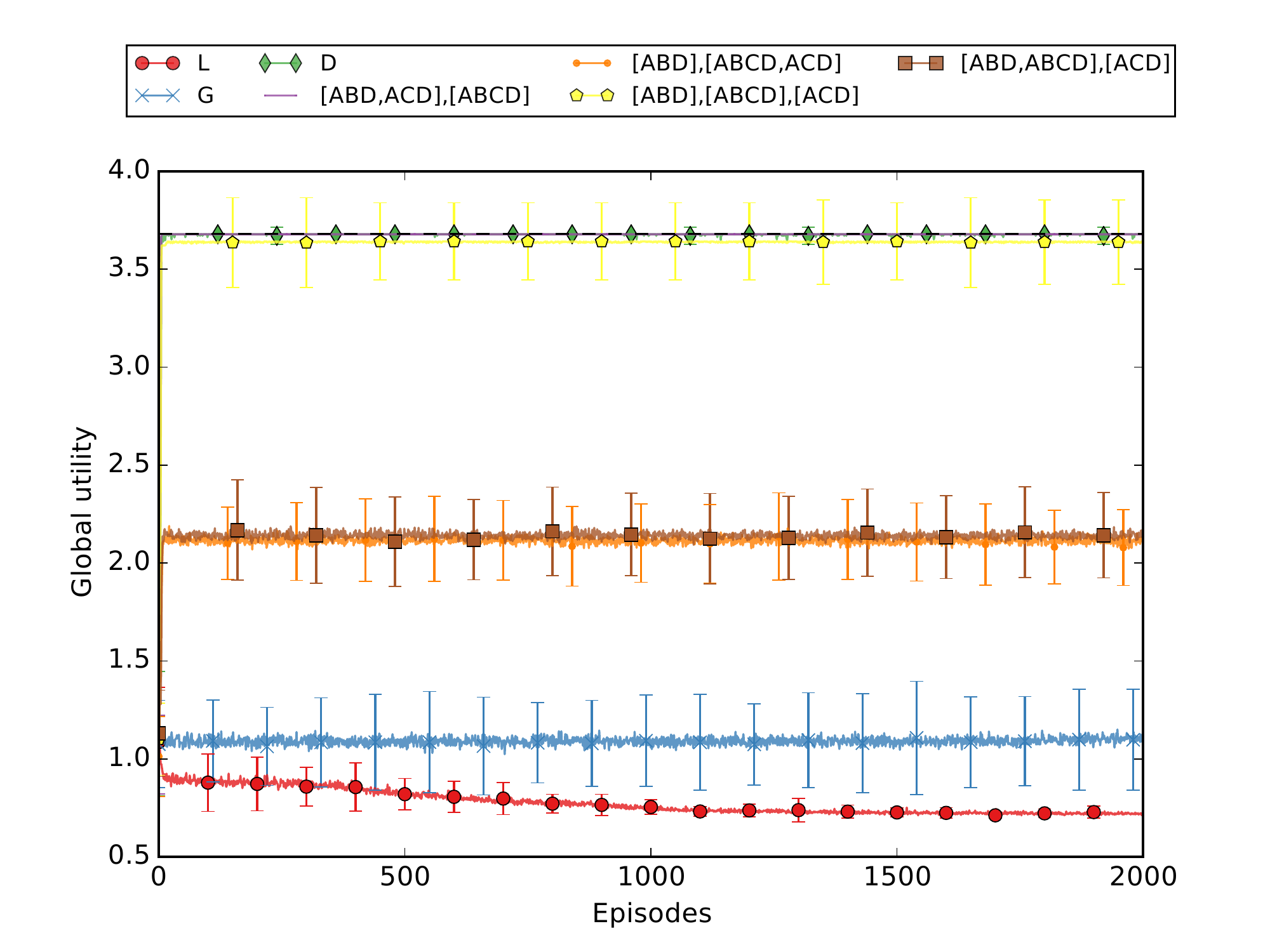}
  \caption{RND with TLD local utility, 50 agents.}
  \label{fig:braessTLD}
\end{figure}

Figure~\ref{fig:braessTLD} presents the obtained results. We notice that in all cases the convergence takes place very fast, however the quality of the solutions vary from one setting to another. The difference rewards approach, $D$, manages again to achieve optimum performance. As expected, the RA configuration $[ABD,ACD],[ABCD]$ also performs optimally, as it explicitly encourages the congestion of the path $ABCD$. $[ABD],[ABCD],[ACD]$ comes close to the optimum performance, however the agents will not always overcrowd path $ABCD$ in this case. The next two configurations, $[ABD,ABCD],[ACD]$ and $[ABD],[ABCD,ACD]$, still perform better than $L$ and $G$, although they encourage a suboptimal behaviour (overcrowd path $ACD$ or $ABD$ respectively). Notice that in comparison to the results in Figure~\ref{fig:braessBPD},  $[ABD,ACD],[ABCD]$ goes from being the worst performing setting to one of the best. This emphasizes how important it is to have insights on what the desired collective behaviour of the agents is, in order to provide well performing abstract groupings. 
\section{Discussion and Conclusions}
\label{sec:conc}
The contribution of this work is two-fold.  Firstly, we introduce a new congestion problem, the Road Network Domain, in which the resources are no longer independent, and the selection of one path influences the load in other parts of the network. Secondly, we provide a thorough analysis of the resource abstraction approach for resource selection congestion problems, together with its limitations and clear guidelines on how to best create the abstract groups according to the desired outcome.

The Road Network Domain presents a novel challenge for resource selection congestion problems, introducing the realistic aspect of interconnected resources as we often find in real-word application such as: electricity grids or traffic networks. We note that the network topology used here is a small one, yet sufficient to illustrate the additional challenge, and that more research is necessary in order to evaluate scenarios that closely model real-world situations. A next possible step would be to investigate the scalability of all the methods, when dealing with a larger and more complex network, especially when considering the challenge of tuning the resource abstraction approach.

While resource abstraction seems to provide a strong method of guiding agents towards an `overcrowd one' behaviour (namely the smallest sized group), it fails when the optimal configuration can no longer be expressed in these terms. Additionally, we consider this method to have a limited applicability in real-world domains, as it requires information regarding each composing resource (capacity, weight, consumption), as well as some intuition on the problem's utility function. 

We have also shown that the difference rewards approach achieves a high performance in all the tested scenarios, managing to capture in the reward signal the necessary information to allow the agents to coordinate indirectly, even in more complex scenarios as the RND. 
 

\section*{Acknowledgments}
This work is supported by Flanders Innovation \& Entrepreneurship (VLAIO), SBO project 140047: Stable MultI-agent LEarnIng for neTworks (SMILE-IT). We are grateful for all the helpful comments and discussions with our VUB AI Lab colleagues at each stage of this work.

\bibliographystyle{abbrv}
\bibliography{references}  
%
\end{document}